\documentclass[showpacs,twocolumn]{revtex4}

\usepackage{graphicx}
\usepackage{dcolumn}
\usepackage{amsmath}
\usepackage{color}

\makeatletter
\def\btt#1{\texttt{\@backslashchar#1}}
\DeclareRobustCommand\bblash{\btt{\@backslashchar}} \makeatother

\begin{document}
\title{Quintessence background for $5D$ Einstein-Gauss-Bonnet black holes}
\author{Sushant G. Ghosh $^{a \;b}$} \email{sghosh2@jmi.ac.in, sgghosh@gmail.com}
\author{Muhammed Amir $^{b}$} \email{amirctp12@gmail.com}
\author{Sunil D. Maharaj $^{a}$}\email{maharaj@ukzn.ac.za}
\affiliation{$^{a}$ Astrophysics and Cosmology
Research Unit, School of Mathematics, Statistics and Computer Science, University of
KwaZulu-Natal, Private Bag 54001, Durban 4000, South Africa}
\affiliation{$^{b}$ Centre for Theoretical Physics, Jamia Millia
Islamia, New Delhi 110025, India}
\date{\today}

\begin{abstract}
As we know that the Lovelock theory is an extension of the general relativity to the 
higher-dimensions, in this theory the first and the second order terms correspond to the general relativity 
and the Einstein-Gauss-Bonnet gravity, respectively. We obtain a $5D$ black hole solution in 
Einstein-Gauss-Bonnet gravity surrounded by the quintessence matter, and also analyze their thermodynamical 
properties. Owing to the quintessence corrected black hole, the thermodynamic quantities have also been 
corrected except for the black hole entropy, and phase transition is achievable. The phase transition for 
the thermodynamic stability, is characterized by a discontinuity in the specific heat at $r=r_C$, with 
the stable (unstable) branch for $r < (>) r_C$. 
\end{abstract}

\maketitle

\section{INTRODUCTION}
The gravity theory with higher-curvature term, the so-called Lovelock is one of the natural generalization 
of Einstein's general relativity, introduced originally by Lanczos \cite{Lanczos}, and rediscovered by 
David Lovelock \cite{dll}, the action of it contains higher-order curvature terms and that reduces to 
the Einstein-Hilbert action in four-dimensions, and its second-order term is the Gauss-Bonnet invariant. 
The Lovelock theories have some special characteristics, among the larger class of general higher-curvature 
theories, in having field equations involving not more than second derivatives of the metric. 
As higher-dimensional members of Einstein's general relativity family, the Lovelock theories allow us 
to explore several conceptual issues of gravity in a broader setup. Hence, these theories receive  
significant attention, especially when finding black hole solutions. Besides, the theory is known to be 
free of ghosts about other exact backgrounds \cite{bd}. The theory represents a very interesting scenario 
to study how higher-order curvature corrections to the black hole physics substantially changes the 
qualitative features, we know from our experience with black holes in general relativity. Since, 
its inception, steady attention has been devoted to black hole solutions, including their formation, 
stability, and thermodynamics. The spherically symmetric static black hole  solution for the Einstein-Gauss-
Bonnet theory was first obtained by Boulware and Deser \cite{bd}, and later several authors explored exact 
black hole solutions and their thermodynamical properties 
\cite{egb,Cai:2001dz,Cai:2003gr,Ghosh,Sahabandu:2005ma,Cai:2003kt}. The generalization of the Boulware-Desser 
solution has been obtained with a source as a cloud of strings, in Einstein-Gauss-Bonnet gravity 
\cite{som1,hr}, and also in Lovelock gravity \cite{sgr,sgsm,Ghosh:2014pga}.  

An intense activity of studying black hole solutions in Einstein-Gauss-Bonnet theory of gravity is due to 
the fact that, besides theoretical results, cosmological evidence, e.g., dark matter and dark energy. The 
quintessence is a hypothetical form of dark energy postulated as an explanation of the observation for an 
acceleration of the Universe, rather than due to a true cosmological constant. If quintessence  exists all 
over in the Universe; it can also be around a black hole. In this paper, we are interested in a solution to 
the Einstein equations with the assumption of spherical symmetry, with the quintessence  matter obtained by 
Kiselev \cite{Kiselev:2002dx} and was also rigorously analyzed by himself and others 
\cite{Kiselev:2002dx,Ma:2007zze,Fernando:2012ue,Feng:2014yjj,Malakolkalami:2015cza,Hussain:2014fca}. In 
particular, spherically symmetric quintessence black hole solutions \cite{Kiselev:2002dx}, which is extended 
to the  higher dimensions \cite{Chen:2008ra}, to include Narai solutions 
\cite{FERNANDO:2013uxa,Fernando:2014wma}, and also to the charged black holes \cite{Azreg-Ainou:2014lua}. 
The black hole thermodynamics for the quintessence corrected solutions were obtained in 
\cite{Wei:2011za,Thomas:2012zzc,AzregAinou:2012hy,Tharanath:2013jt,Ghaderi:2016dpi} and, quasinormal modes 
of such solutions are also discussed \cite{Zhang:2007nu,Varghese:2008ky,Saleh:2011zz,Tharanath:2014uaa}. The 
generalization of the spherical quintessential solution \cite{Kiselev:2002dx} to the axially symmetric case, 
Kerr-like black hole, was also addressed, recently \cite{Ghosh:2015ovj,Toshmatov:2015npp}. However, the 
solution of the Einstein-Gauss-Bonnet theory surrounded by the quintessence matter is still not explored, 
i.e., the black holes surrounded by the quintessence matter in Einstein-Gauss-Bonnet theory is still unknown. 
It is the purpose of this paper to obtain an exact new five-dimensional ($5D$) spherically symmetric 
black holes solution for the Einstein-Gauss-Bonnet surrounded by the quintessence matter. 
In particular, we explicitly bring out how the effect of a background quintessence matter can alter 
black hole solutions and their thermodynamics. In turn,  we analyze their thermodynamical properties 
and perform a  thermodynamic stability analysis.

The paper is organized as follows. In Sec.~\ref{QM}, we derive a Einstein-Gauss-Bonnet solution to the 
$5D$ spherically symmetric static Einstein's equations surrounded by the quintessence matter. 
In Sec.~\ref{thermo}, we have discussed the thermodynamics of the $5D$ Einstein-Gauss-Bonnet black holes 
surrounded by the quintessence matter. The paper ends with concluding remarks in Sec.~\ref{conclusion}.  

We use units which fix the speed of light and the gravitational constant via $G = c = 1$, and use the metric signature ($-,\;+,\;+,\;+,\;+$).

\section{Quintessence matter surrounding black hole}
\label{QM}
The Lovelock theory is an extension of the general relativity to higher-dimensions. In this theory the first and second order terms correspond to the general relativity and the Einstein-Gauss-Bonnet, respectively. 
The action for  $5D$ Einstein-Gauss-Bonnet gravity with matter field reads:  
\begin{equation}\label{action}
\mathcal{I}_{G}=\frac{1}{2}\int_{\mathcal{M}}d^{5}x\sqrt{-g}\left[  \mathcal{L}_{1} +\alpha \mathcal{L}_{GB}
 \right] + \mathcal{I}_{S},
\end{equation}
with $\kappa_5 =1$. $\mathcal{I}_{S}$ denotes the action associated with matter and $\alpha$ is coupling constant that we assume to be non-negative. The Einstein term is $ \mathcal{L}_1 = R$, and the second-order Gauss-Bonnet term $\mathcal{L}_{GB}$ is
\begin{equation}
\mathcal{L}_{GB}=R_{\mu\nu\gamma\delta}R^{\mu \nu\gamma\delta}-4R_{\mu\nu}R^{\mu\nu}+R^{2}.
\end{equation}
Here, $R_{\mu\nu}$, $R_{\mu\nu\gamma\delta\text{}}$, and $R$  are the Ricci tensor, Riemann tensor, and  Ricci scalar, respectively. The variation of the action with respect to the metric $g_{\mu\nu}$ gives the Einstein-Gauss-Bonnet equations
\begin{equation}\label{ee}
G_{\mu\nu}^{E}+\alpha G_{\mu\nu}^{GB}=T_{\mu\nu}^{S},
\end{equation}
where $G_{\mu\nu}^{E}$ is the Einstein tensor while $G_{\mu\nu}^{GB}$ is given explicitly by \cite{Kastor:2006vw}
\begin{eqnarray}
 G_{\mu\nu}^{GB} & = & 2\;\Big[ -R_{\mu\sigma\kappa\tau}R_{\quad\nu}^{\kappa
\tau\sigma}-2R_{\mu\rho\nu\sigma}R^{\rho\sigma}-2R_{\mu\sigma}R_{\ \nu
}^{\sigma} \nonumber \\ & &  +RR_{\mu\nu}\Big] -\frac{1}{2}\mathcal{L} _{GB}g_{\mu\nu},
\end{eqnarray}
and $T^S_{\mu\nu}$ is the energy-momentum tensor of the matter that we consider as a quintessence matter. We note that the divergence of Einstein-Gauss-Bonnet tensor $G_{\mu \nu}^{GB}$ vanishes. Here, we want to obtain $5D$ static spherically symmetric solutions of Eq.~(\ref{ee}) surrounded by the quintessence matter and investigate its properties. We assume that the metric has the form \cite{Ghosh,Ghosh:2014pga}
\begin{equation}\label{metric}
ds^2 = -f(r) dt^2+ \frac{1}{f(r)} dr^2 + r^2 \tilde{\gamma}_{ij}\; dx^i\; dx^j,
\end{equation}
where $ \tilde{\gamma}_{ij} $ is the metric of a $3D$ constant curvature space $k = -1,\; 0,\;$ or $1$. In this paper, we shall restrict to $k = 1$.  Using this metric \textit{ansatz}, the Einstein-Gauss-Bonnet Eq.~(\ref{ee}) reduces to 
\begin{eqnarray}\label{EMT}
T^t_t &=& T^r_r= \frac{3}{2r^2}\left[r f' + 3 (f-1)\right] - \frac{6 \alpha}{r^3} \left[(f-1)f'\right], 
\nonumber \\ 
T^{\theta}_{\theta} &=& T^{\phi}_{\phi} = T^{\psi}_{\psi} = \frac{1}{2r^2} \left[r^2f''+4rf'+2(f-1)\right]
\nonumber \\
&& -\frac{\alpha}{r^2} \left[2 (f-1)f''+2f'^2\right].
\end{eqnarray}
The energy-momentum tensor of the quintessence matter (see Ref. \cite{Kiselev:2002dx} for further details) gets modified to
\begin{eqnarray}
T^t_t &=& \rho(r), 
\nonumber \\
T^b_a &=& \rho(r)  \beta \left[-(1+ 4 B(r)) \frac{r_ar^b}{r_nr^n}+ B(r) \delta_a^b \right],
\end{eqnarray}
where $B(r)$ is a quintessential parameter
\begin{equation}
\langle  T^b_a \rangle = \rho(r) \frac{\beta}{4} \delta^b_a = -p(r) \delta^b_a , 
\end{equation}
and $$\langle r_ar^b = \frac{1}{4} r_nr^n \rangle. $$ Thus, we have the equation of state of the form
\begin{equation}
\label{eos}
p=\omega \rho, \quad \omega=\frac{1}{4} \beta ,
\end{equation}
where for the quintessential matter $-1 <\omega<0 $, which implies $-4 < \beta < 0$ within this set up, the parameter $B$ of the energy-momentum tensor reads \cite{Chen:2008ra}
\begin{equation}
B = -\frac{4\omega+1}{\omega}.
\end{equation}
Hence, the energy-momentum tensor for a quintessence matter takes the form
\begin{eqnarray}\label{EMTQ}
T^t_t &=& T^r_r= \rho, 
\nonumber \\
T^{\theta}_{\theta} &=& T^{\phi}_{\phi} = T^{\psi}_{\psi} = - \frac{1}{3} \rho \left(4\omega +1\right),
\end{eqnarray}
where $\rho$ is the proper density of the quintessence matter. The range of the parameters $\omega$ and $\beta$ gets modified in $5$ dimensions. 
\begin{figure*}
\begin{tabular}{c c c c}
\includegraphics[width=0.5\linewidth]{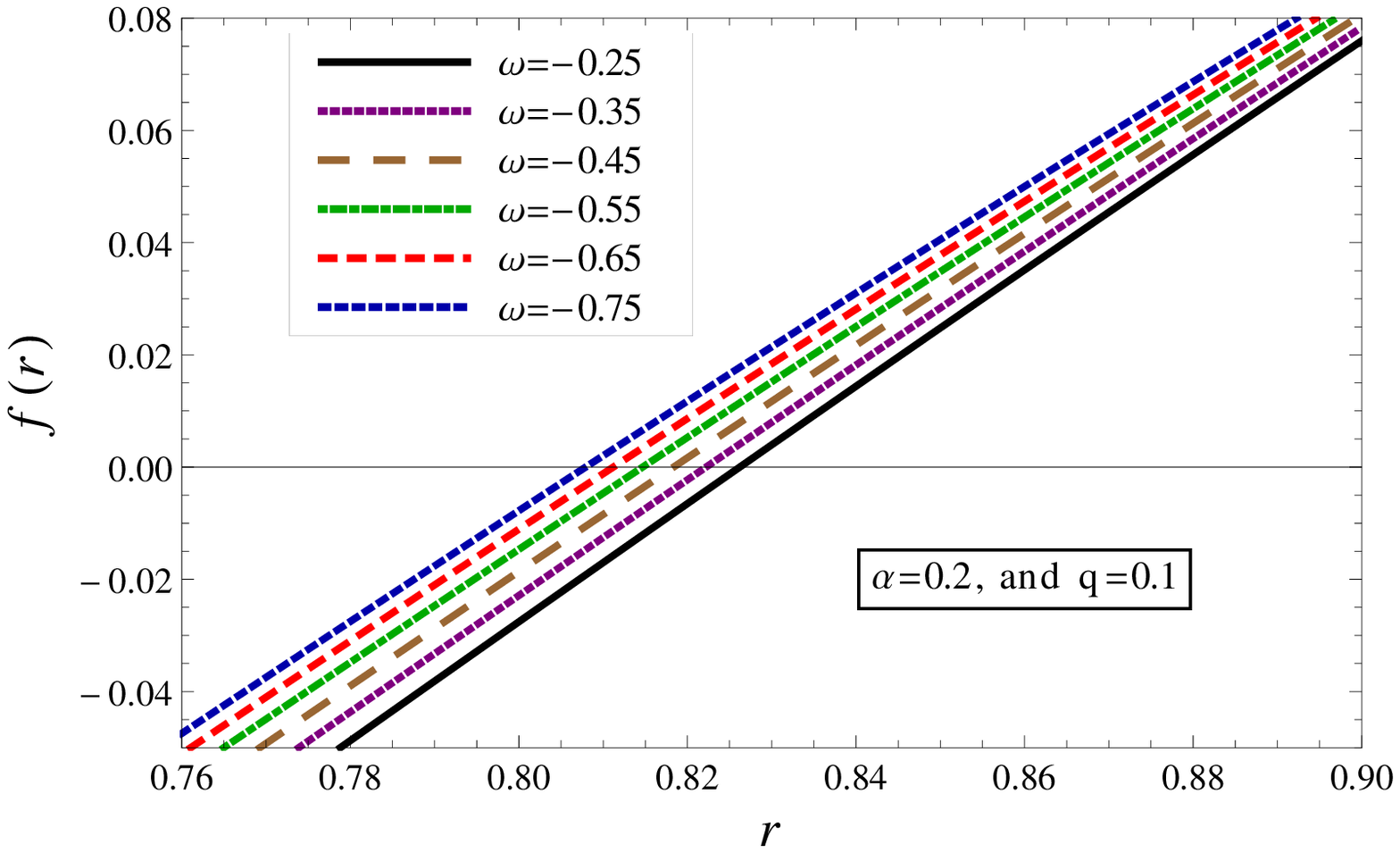}
\includegraphics[width=0.5\linewidth]{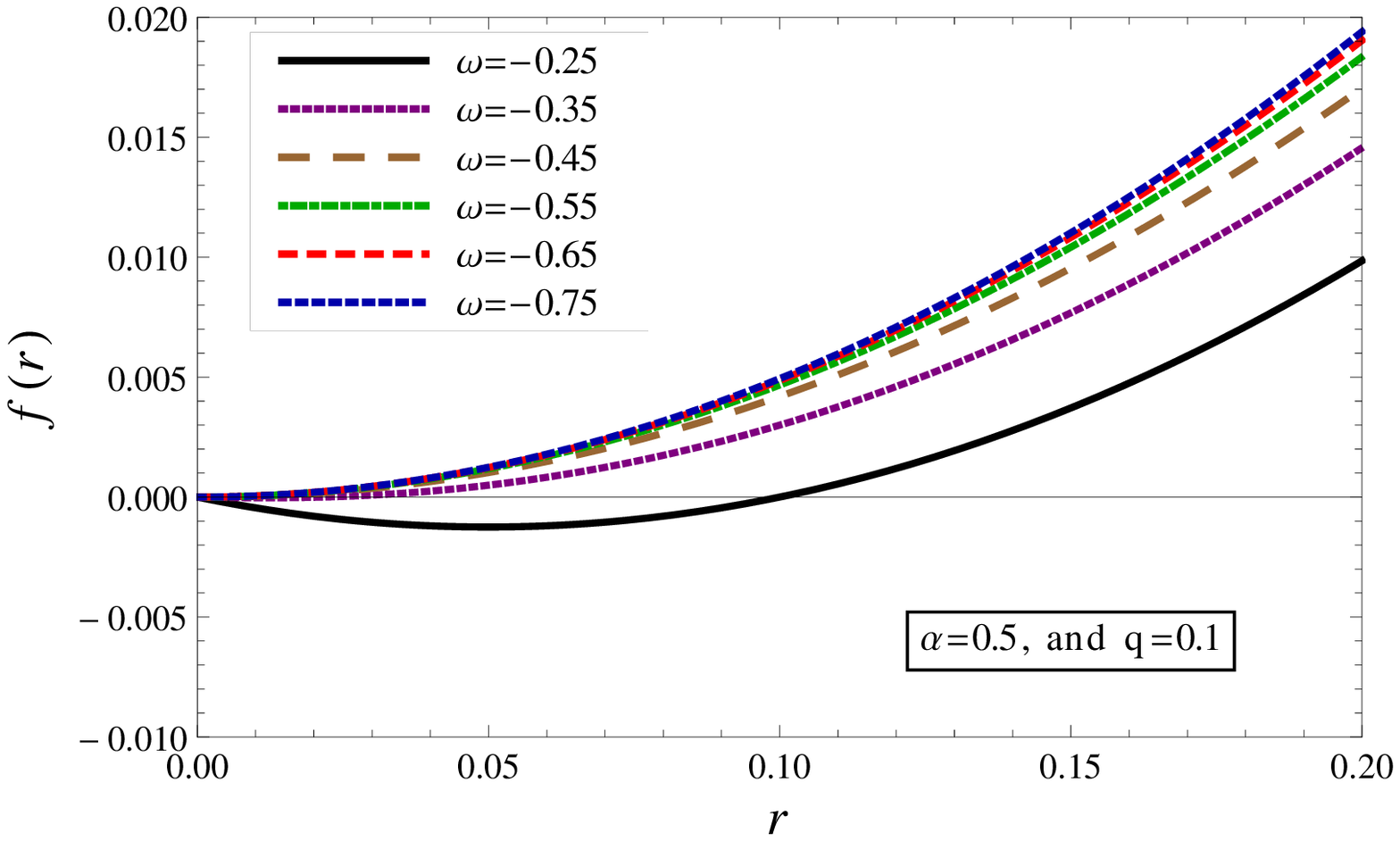}
\\
\includegraphics[width=0.5\linewidth]{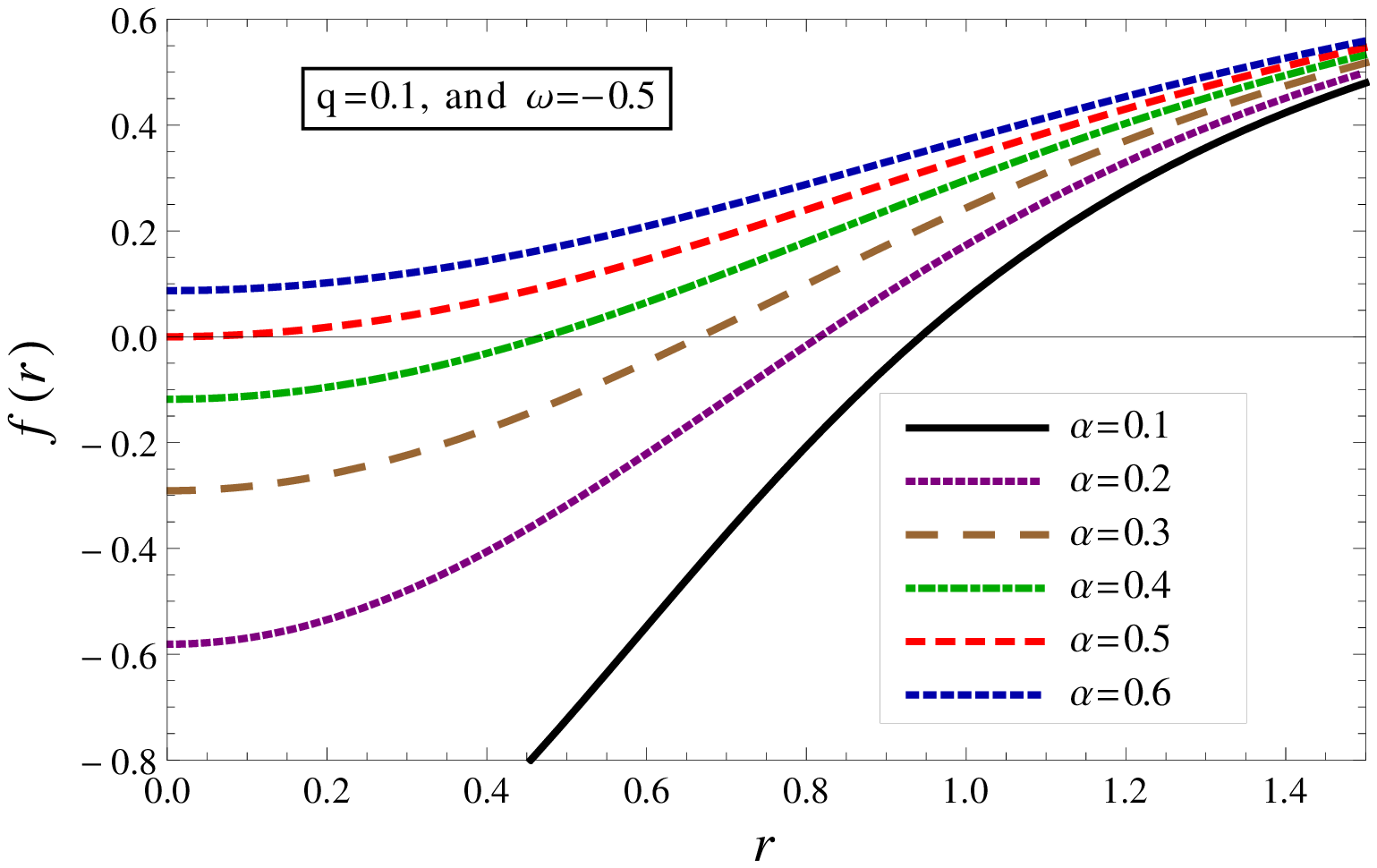}
\includegraphics[width=0.5\linewidth]{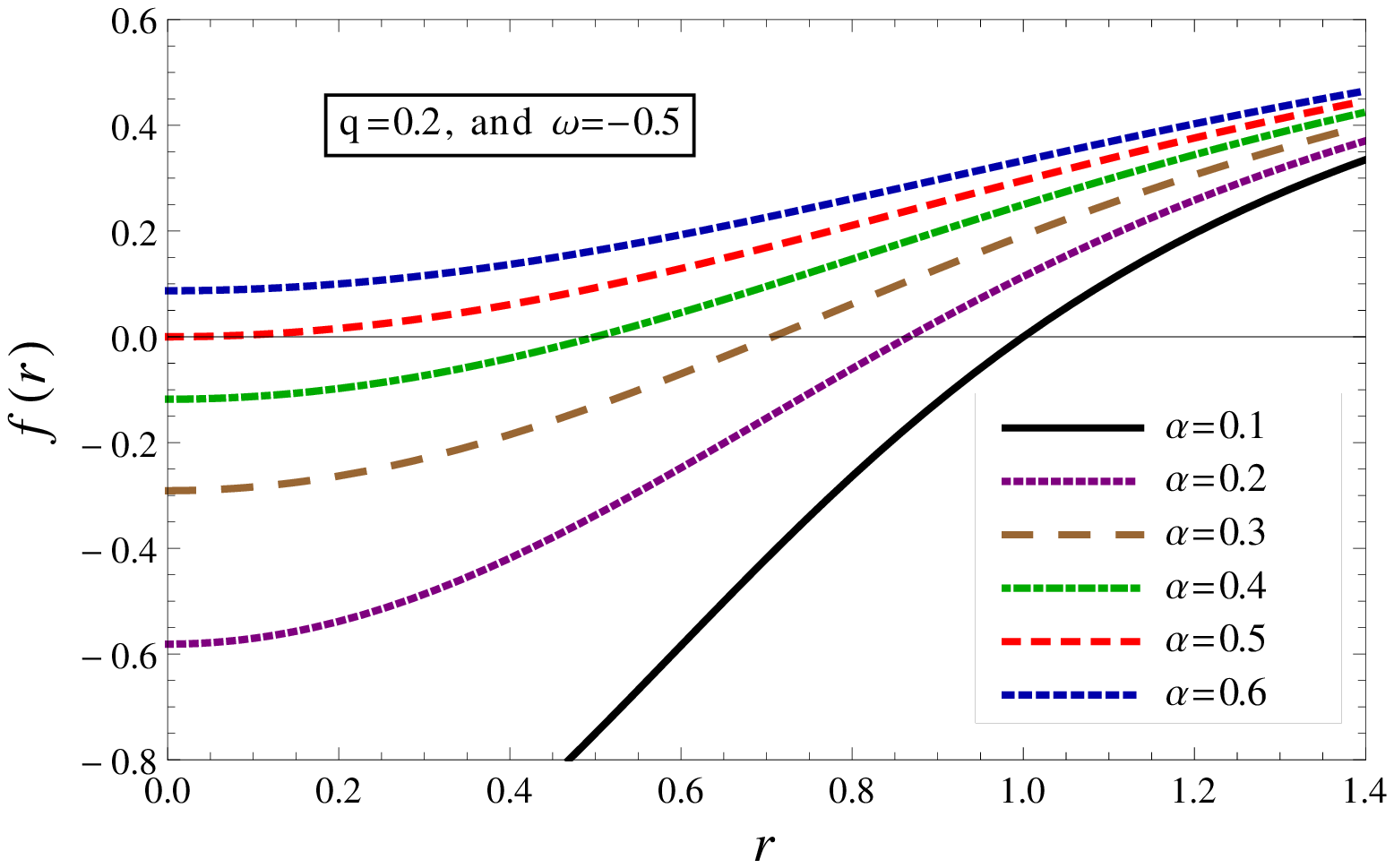}
\end{tabular}
\caption{\label{f5d} Plot of metric function $f(r)$ vs $r$ for the $5D$ Einstein-Gauss-Bonnet black hole 
surrounded by the quintessence matter.}
\end{figure*}

The Gauss-Bonnet term is the only  possibility for the leading correction to Einstein general relativity.
The static spherically symmetric black hole solution of Einstein action, modified by the 
Einstein-Gauss-Bonnet term, was first obtained by Boulware and Deser \cite{bd} and demonstrated that the only 
spherically symmetric solution to the Einstein-Gauss-Bonnet theory is Schwarzschild type solution. 
Later, Wiltshire \cite{Wiltshire:1988uq} included the Maxwell field to the Einstein-Gauss-Bonnet action, 
found the charged black hole in this theory which was a generalization of the Reissner-Nordstr{\"o}m black 
hole. In general, it is difficult to find solution of Einstein-Gauss-Bonnet field equations (\ref{ee}) 
with an equation of state. Here, we shall find a black hole solution surrounded by the quintessence 
matter with the equation of state (\ref{eos}). Making use of Eqs.~(\ref{EMT}) and (\ref{EMTQ}), we deduce 
the master equation for the Einstein-Gauss-Bonnet gravity as
\begin{eqnarray}\label{master}
&& \Big[r^2 f''(r) + (5+4 \omega)rf'(r)+2(2+4\omega)(f(r)-1)\Big]r
\nonumber \\ 
&& -\alpha\Big[4rf''(r)(f(r)-1)\nonumber \\
&& +4[rf'(r) +(1+ 4 \omega)(f(r)-1)f'(r)]\Big] =0.
\end{eqnarray}
In general, Eq.~(\ref{master}) has one real and two complex solutions. It may have three real solutions as well under some conditions. Here, we consider only the real solution. Interestingly, Eq.~(\ref{master}), for the Einstein-Gauss-Bonnet case, admits a general solution 
\begin{equation} \label{sol:egb}
f_{\pm}(r) = 1+\frac{r^{2}}{4{\alpha}}\left(1\pm
\sqrt{1+\frac{8\alpha M}{r^{4}} +\frac{8\alpha q}{r^{4\omega+4}}}\right),
\end{equation}
by appropriately relating $M$ and $q$ with integrating constants $c_1$ and $c_2$ \cite{hr}. Equation 
(\ref{sol:egb}) is an exact solution of the field equation (\ref{master}) for an equation of state (\ref{eos}) 
which in case of no quintessence $\omega=0$ reduces to Boulware and Deser \cite{bd} Gauss-Bonnet black hole 
solution, and for $\omega=1/2$ and $q= - 4Q^2/3$ to a solution mathematically similar to the 
the charged Gauss-Bonnet black hole due to Wiltshire \cite{Wiltshire:1988uq}. When $\omega=-1, q=\Lambda/3$, 
the Eq.~(\ref{sol:egb}) corresponds to Gauss-Bonnet de Sitter solution. In the limit $\alpha \rightarrow 0$, the negative branch of the solution (\ref{sol:egb}) reduces to the general relativity solution. To study the general structure of the solution (\ref{sol:egb}), we take the limit $r\rightarrow\infty$ or $M=q=0$ in solution (\ref{sol:egb}) to obtain
\begin{equation}
\lim_{r\rightarrow \infty} f_+(r)= 1+\frac{r^2}{2\alpha},\;\;\; \lim_{r\rightarrow\infty} f_-(r)= 1,
\end{equation}
this means the plus ($+$) branch of the solution (\ref{sol:egb}) is asymptotically de Sitter (anti-de Sitter) depending on the sign of $\alpha$ $(\pm)$, whereas the minus branch of the solution (\ref{sol:egb}) is asymptotically flat. In the large $r$ limit (or $\alpha \rightarrow 0$), Eq.~(\ref{sol:egb}) reduces to the $5D$ Schwarzschild solution surrounded by the quintessence matter. Thus, the negative branch solution (\ref{sol:egb}) is well behaved and it represents short distance correction to $5D$ black hole solution of general relativity. In similar way, when $M=0$, the solution (\ref{sol:egb}) takes the form
\begin{equation} \label{sol:egb2}
f_{\pm}(r) = 1+\frac{r^{2}}{4{\alpha}}\left(1\pm \sqrt{1 +\frac{8\alpha q}{r^{4\omega+4}}}\right).
\end{equation}
Obviously, by proper choice of the functions $M$ and $q$, and parameter $\omega$, one can generate many other known solutions. The above solutions include most of the known spherically symmetric solutions  of the Einstein-Gauss-Bonnet  field equations (\ref{ee}).

\section{Thermodynamics}
\label{thermo}
\begin{figure*}
\begin{tabular}{ c c c c}
\includegraphics[width=0.5\linewidth]{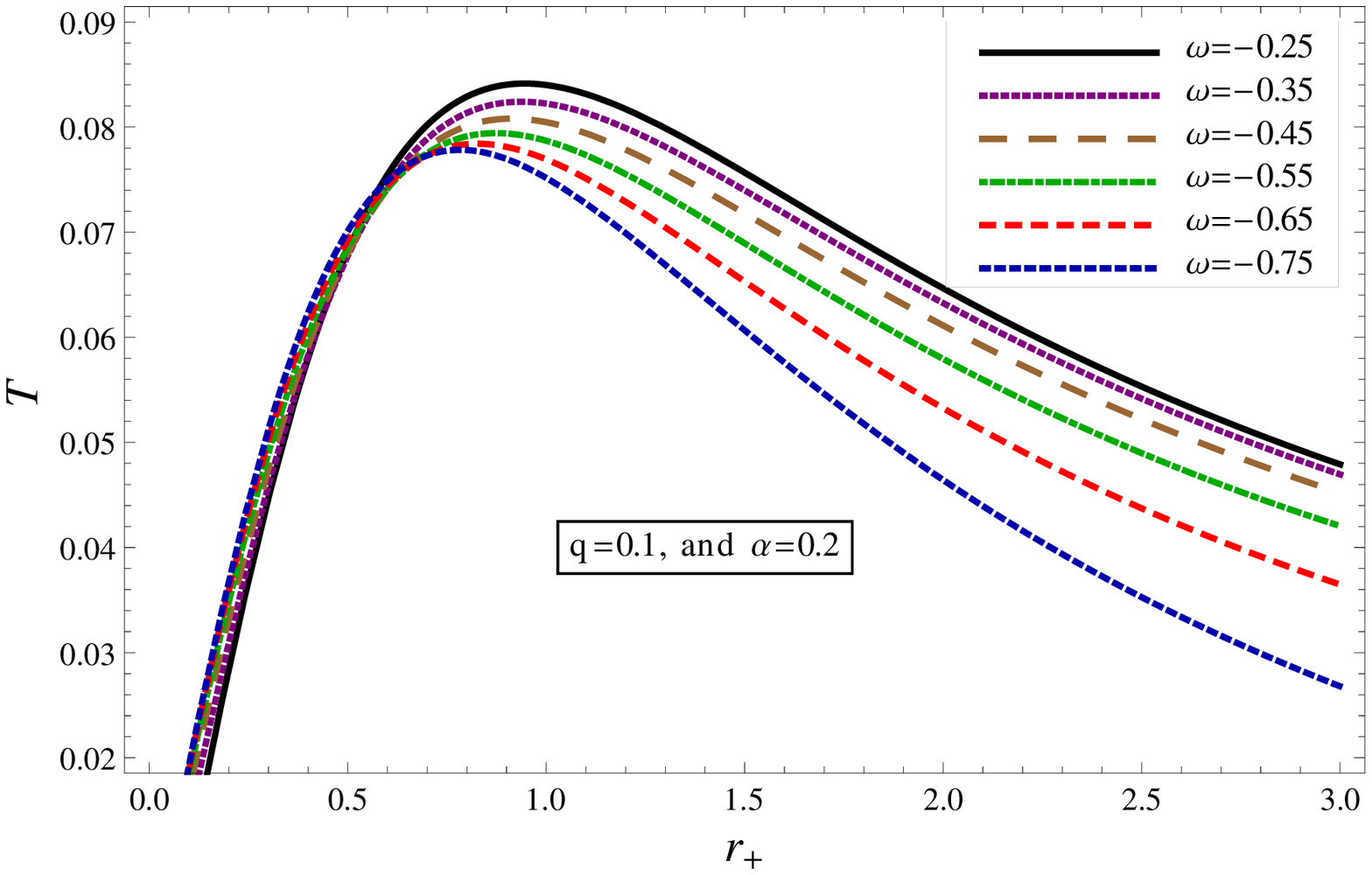}
\includegraphics[width=0.5\linewidth]{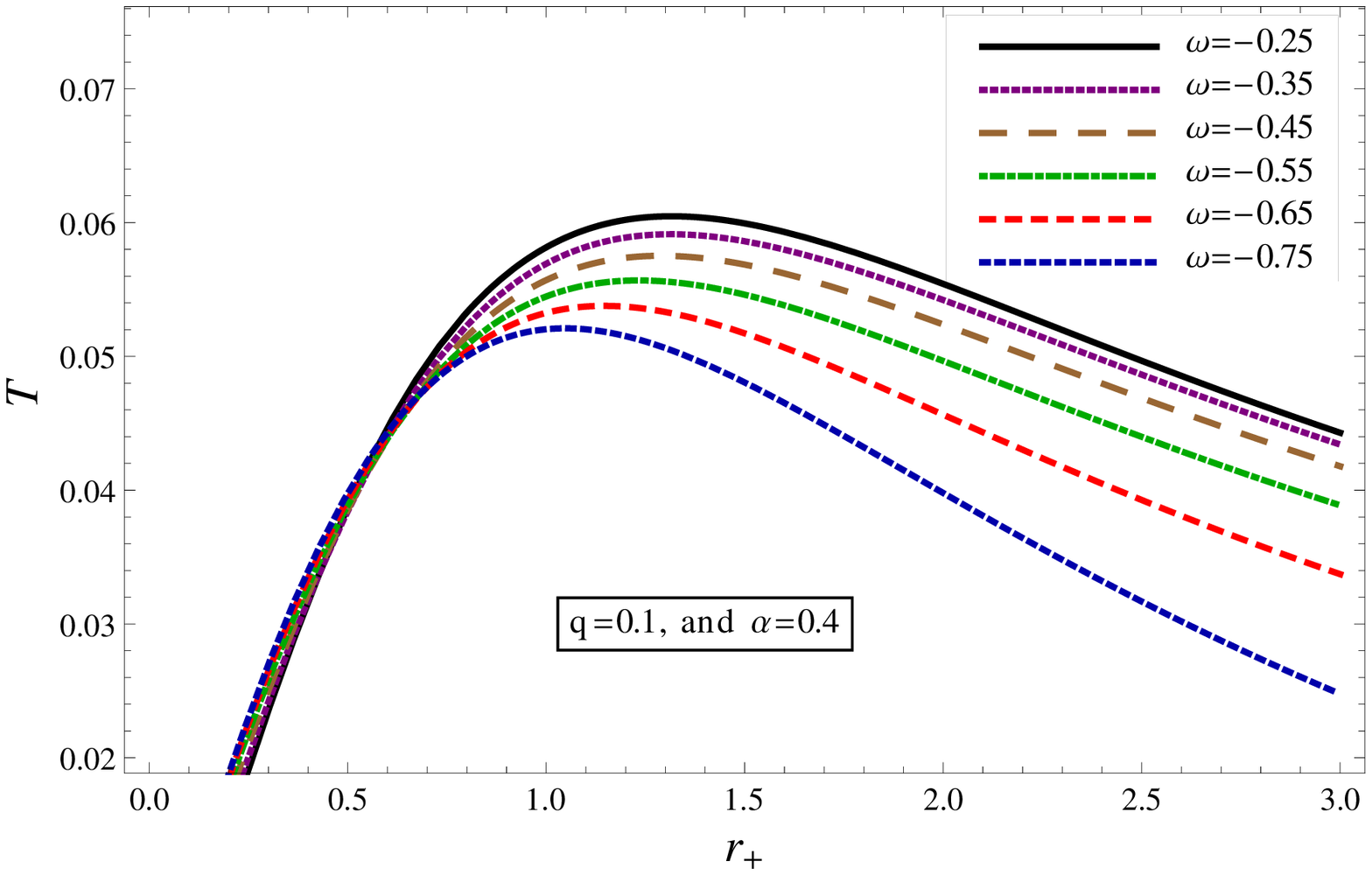}
\\
\includegraphics[width=0.5\linewidth]{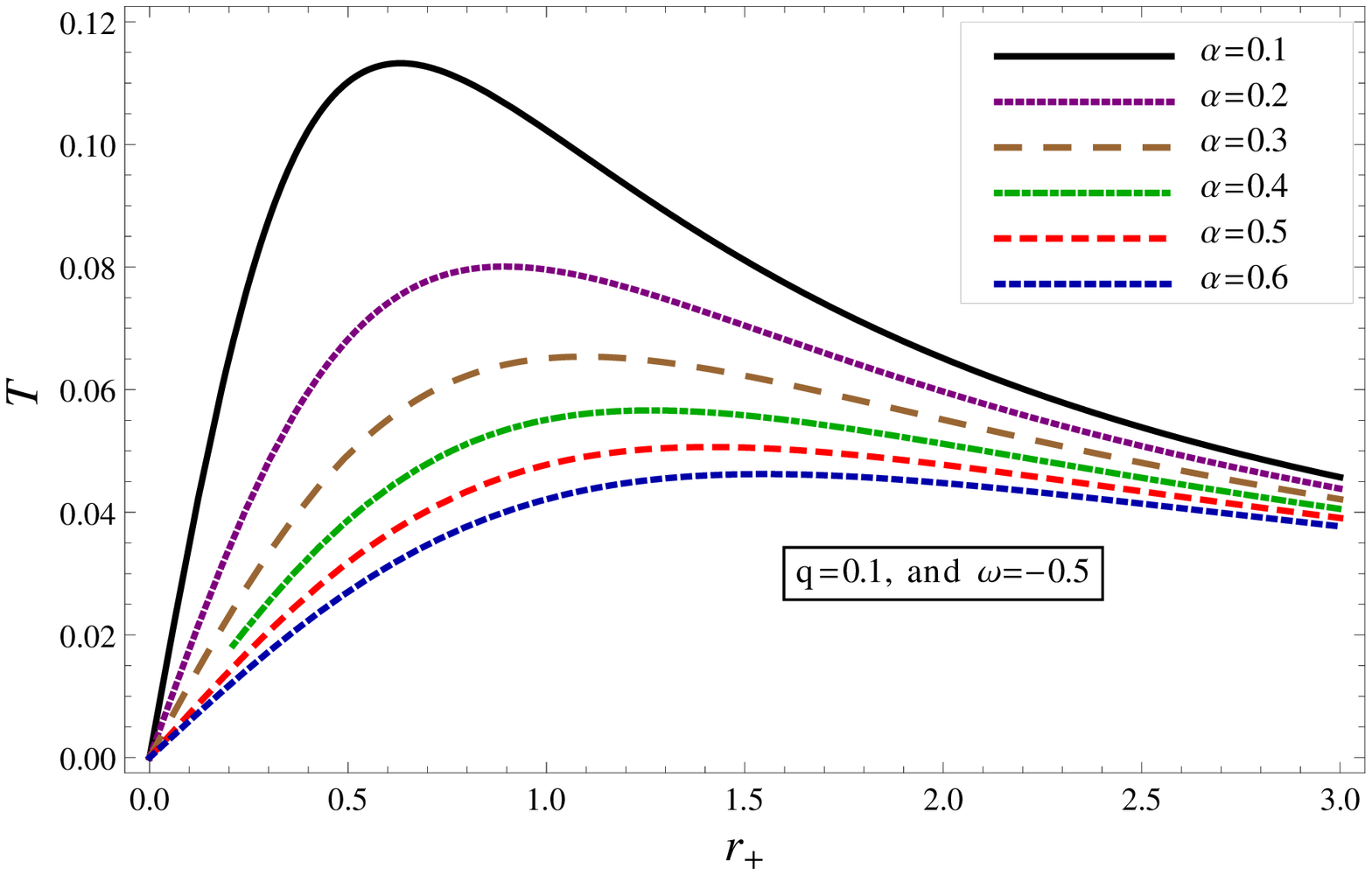}
\includegraphics[width=0.5\linewidth]{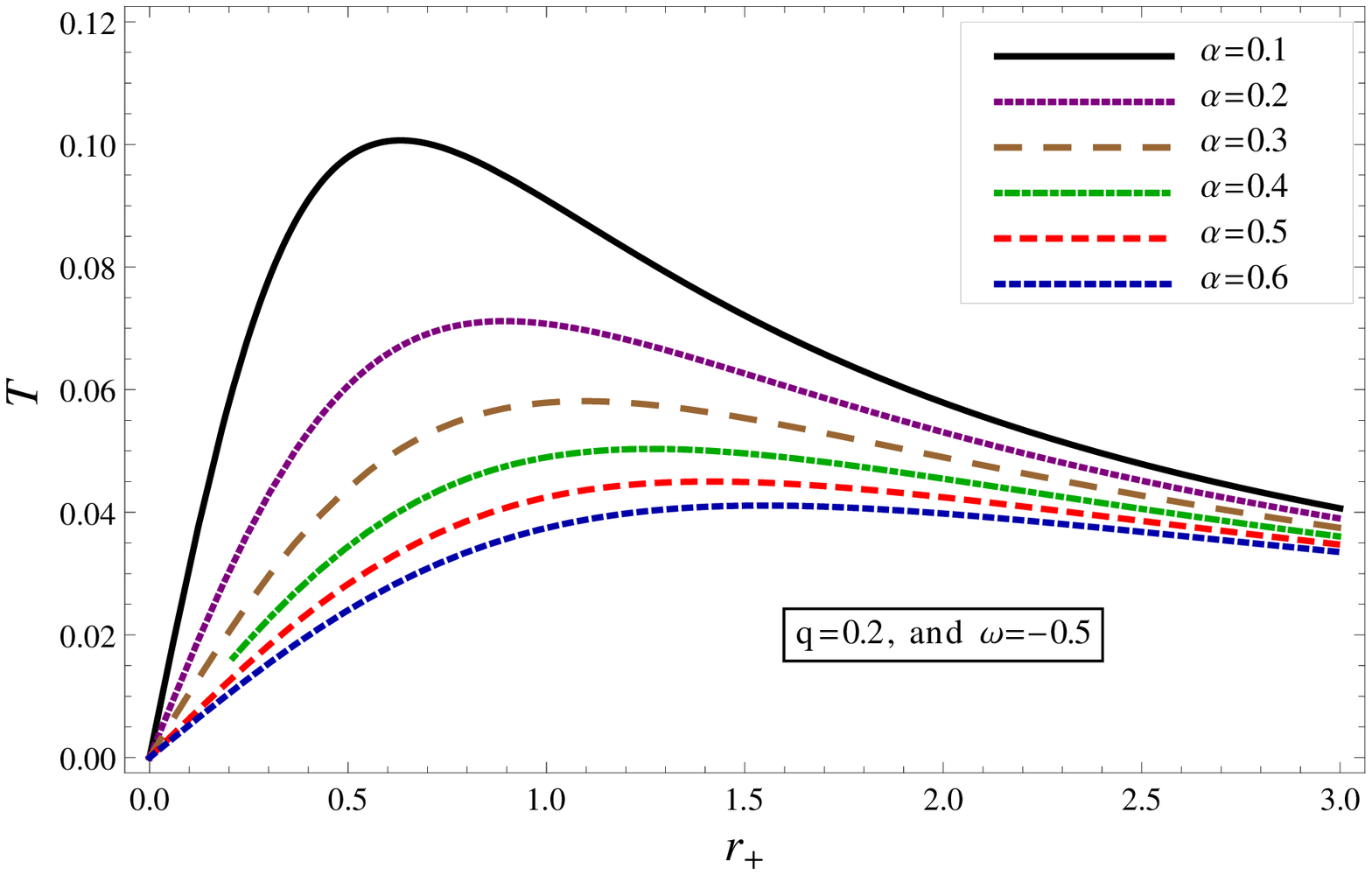}
\end{tabular}
\caption{\label{fig:egb:T} The Hawking temperature ($ T_+^{\mbox{EGB}} $ ) vs horizon radius $r_+$ 
for different values of $\alpha$ $\omega$ with fixed $q=0.1$ for the $5D$ Einstein-Gauss-Bonnet black hole
surrounded by the quintessence matter.}
\end{figure*}

In this section, we shall discuss the thermodynamical properties of $5D$ quintessential black hole within Einstein-Gauss-Bonnet framework. Henceforth, we shall restrict ourselves to the negative branch of the solution (\ref{sol:egb}). By definition of a horizon, the value of $r=r_+ $ is an event horizon when $f(r_+)=0$. This is shown in Fig.~\ref{f5d}, by plotting $f(r)$ as a function of $r$. It is interesting to note that the black holes admit only one horizon and the radius of the horizon increases with the value of the quintessence matter parameter $\omega$.
\begin{figure*}
\begin{tabular}{c c}
\includegraphics[width=0.5\linewidth]{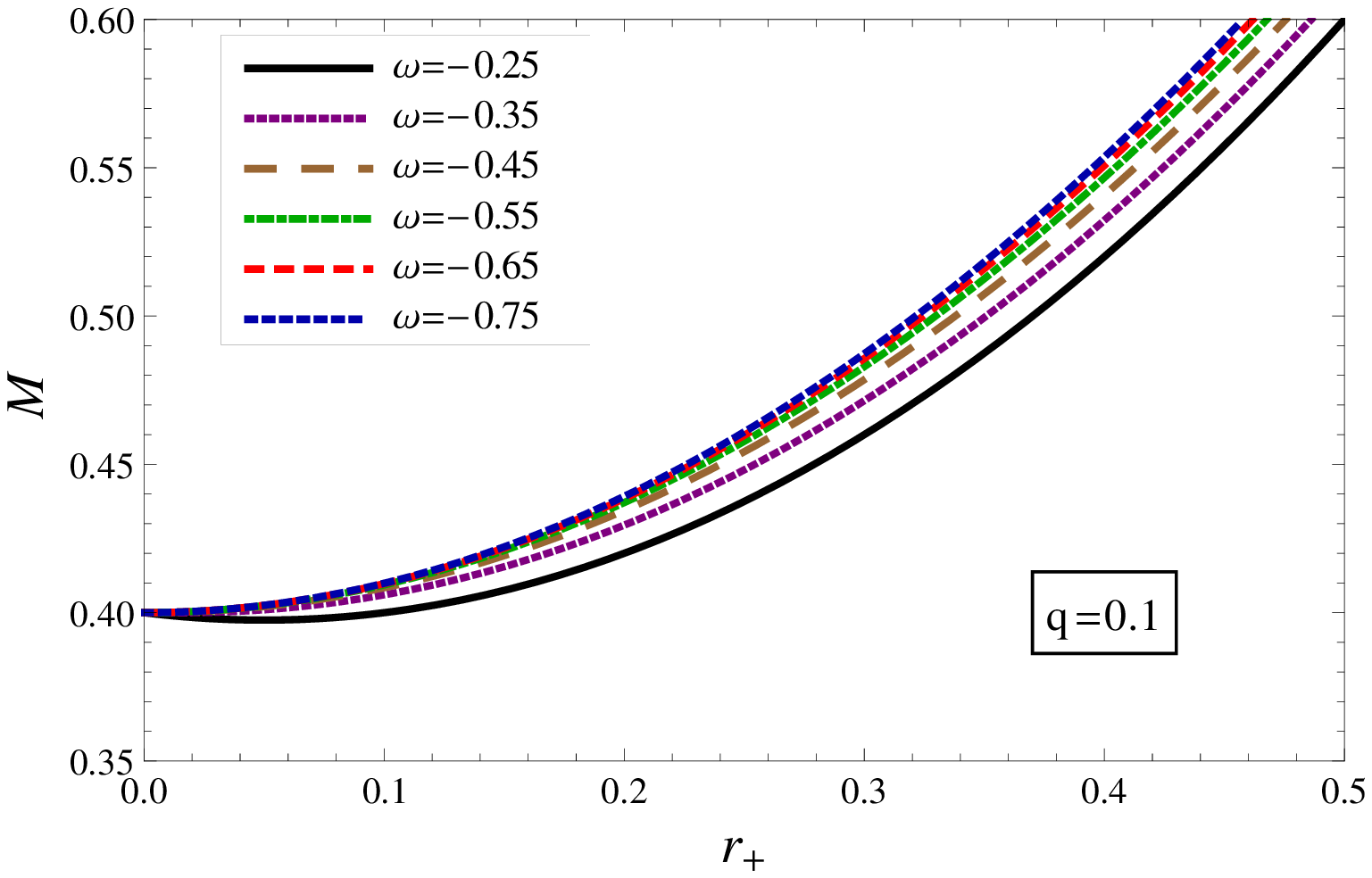}
\includegraphics[width=0.5\linewidth]{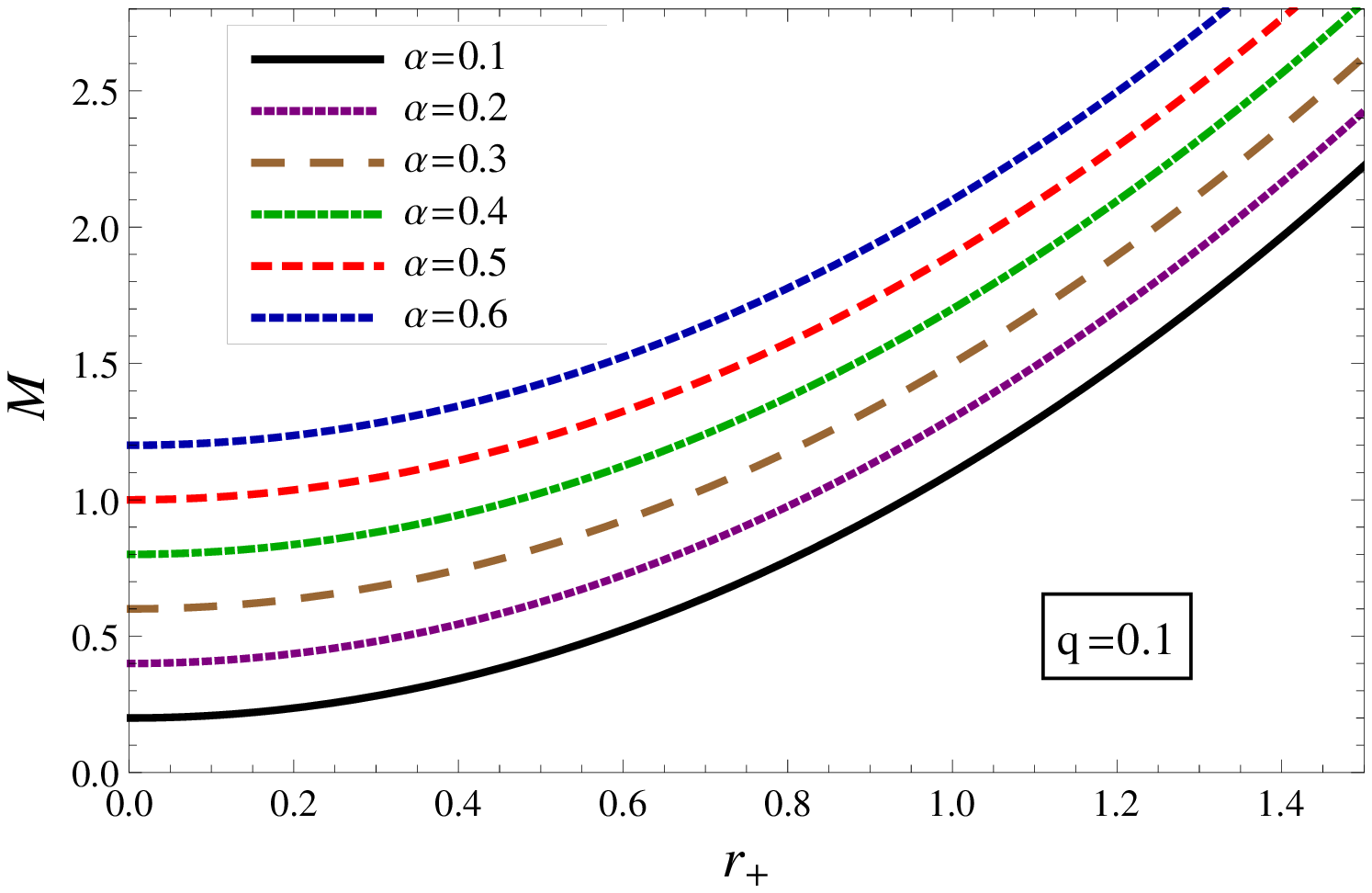}
\end{tabular}
\caption{\label{M_vs_r} Plot showing the behavior of mass $M_+^{\mbox{EGB}}$ vs horizon radius $r_+$ for $5D$ Einstein-Gauss-Bonnet black hole surrounded by the quintessence matter.}
\end{figure*}
Next, we explore the thermodynamics of the black hole solution (\ref{sol:egb}) surrounded by the quintessence matter in Einstein-Gauss-Bonnet framework. The Einstein-Gauss-Bonnet black holes surrounded by the quintessence matter are characterized by their mass $(M)$ and a quintessence matter parameter ($\omega$). From Eq.~(\ref{sol:egb}), the mass of the black hole is obtained in terms of the horizon radius ($r_+$):
\begin{eqnarray}\label{M1}
M_{+}^{\mbox{EGB}} =  r_{+}^2\left(1-\frac{q}{r_{+}^{4\omega +2}} +\frac{2\alpha}{r_{+}^2}\right). 
\end{eqnarray}
The mass of the black hole is plotted in Fig.~\ref{M_vs_r} for various values of parameters $\omega$ and $\alpha$, which shows an increase in the black hole mass with horizon radius $r_+$. To discuss the thermodynamics of the metric (\ref{metric}) with function (\ref{sol:egb}), we start with the Hawking temperature. The Hawking temperature associated with the black hole is defined by $T=\kappa/2\pi$, where $\kappa$ is the surface gravity defined by \cite{Sahabandu:2005ma,Ghosh:2014pga},
\begin{equation}
\kappa^{2}=-\frac{1}{4}g^{tt}g^{ij}g_{tt,i}\;g_{tt,j}.
\end{equation}
Hence, the Hawking temperature for the Einstein-Gauss-Bonnet black hole surrounded by the quintessence matter can be calculated as
\begin{eqnarray}\label{temp1}
T_+^{\mbox{EGB}}  &=& \frac{1}{2 \pi r_{+}} \left[\frac{1 +\frac{2q\omega}{r_{+}^{4\omega+2}}}{1+\frac{4\alpha}{r_{+}^2}}\right].
\end{eqnarray}
Note that the factor in the numerator of Eq.~(\ref{temp1}) modifies the Gauss-Bonnet black hole temperature \cite{Sahabandu:2005ma,Ghosh:2014pga}, and taking the limit $q\rightarrow0$, we recover the Gauss-Bonnet black hole temperature as
\begin{equation}
T_{+}^{\mbox{EGB}} =\frac{1}{2\pi r_+}\left[\frac{r_{+}^2}{r_{+}^2+4\alpha}\right],
\end{equation} 
and when $\alpha \rightarrow 0$, it becomes the temperature given by $T_{+} =\frac{1}{2\pi r_+}$. It is interesting to note that for a particular radius of the horizon, the Hawking temperature vanishes. The Hawking temperature diverges in general relativity as $r_+ \rightarrow 0$. However, in Einstein-Gauss-Bonnet it remains finite as shown in Fig.~\ref{fig:egb:T}. Also, when $q \neq 0$ and $\alpha \neq 0$, the Hawking temperature has a peak which decreases and shifts as $\alpha$ increases or $q$ increases (cf., Fig.~\ref{fig:egb:T}). 
\begin{figure*}
\begin{tabular}{ c c c c}
\includegraphics[width=0.5\linewidth]{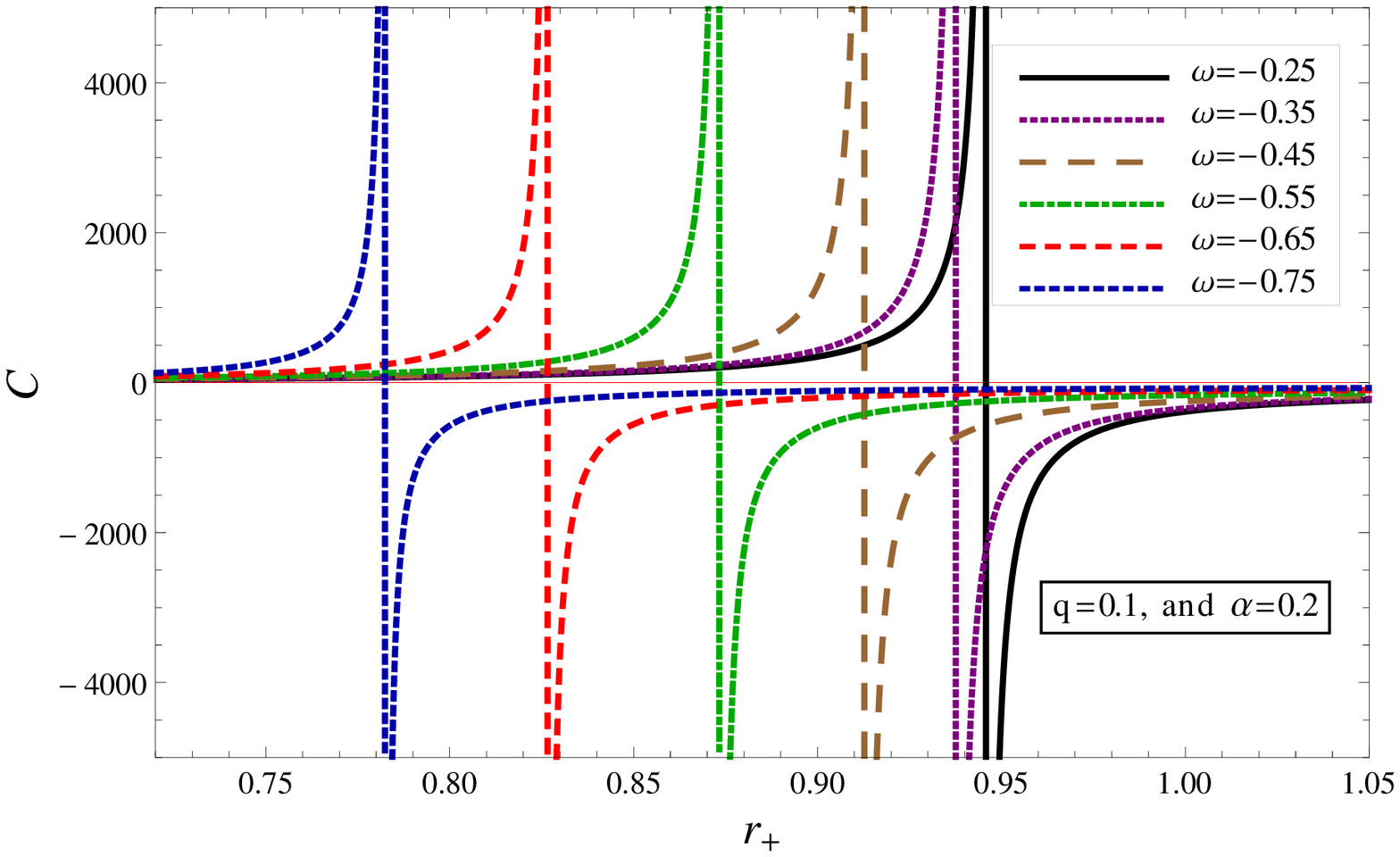}
\includegraphics[width=0.5\linewidth]{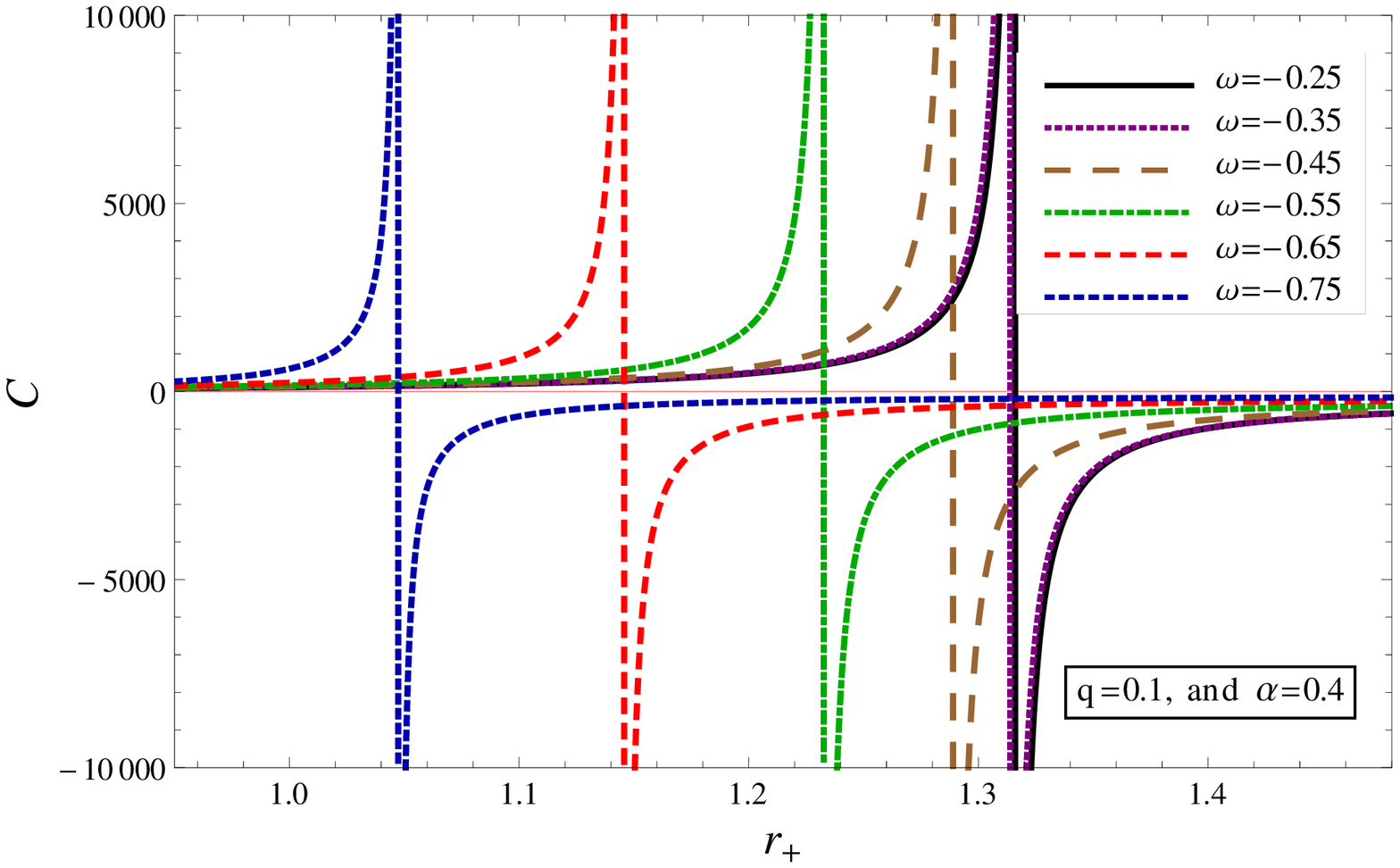}
\\
\includegraphics[width=0.5\linewidth]{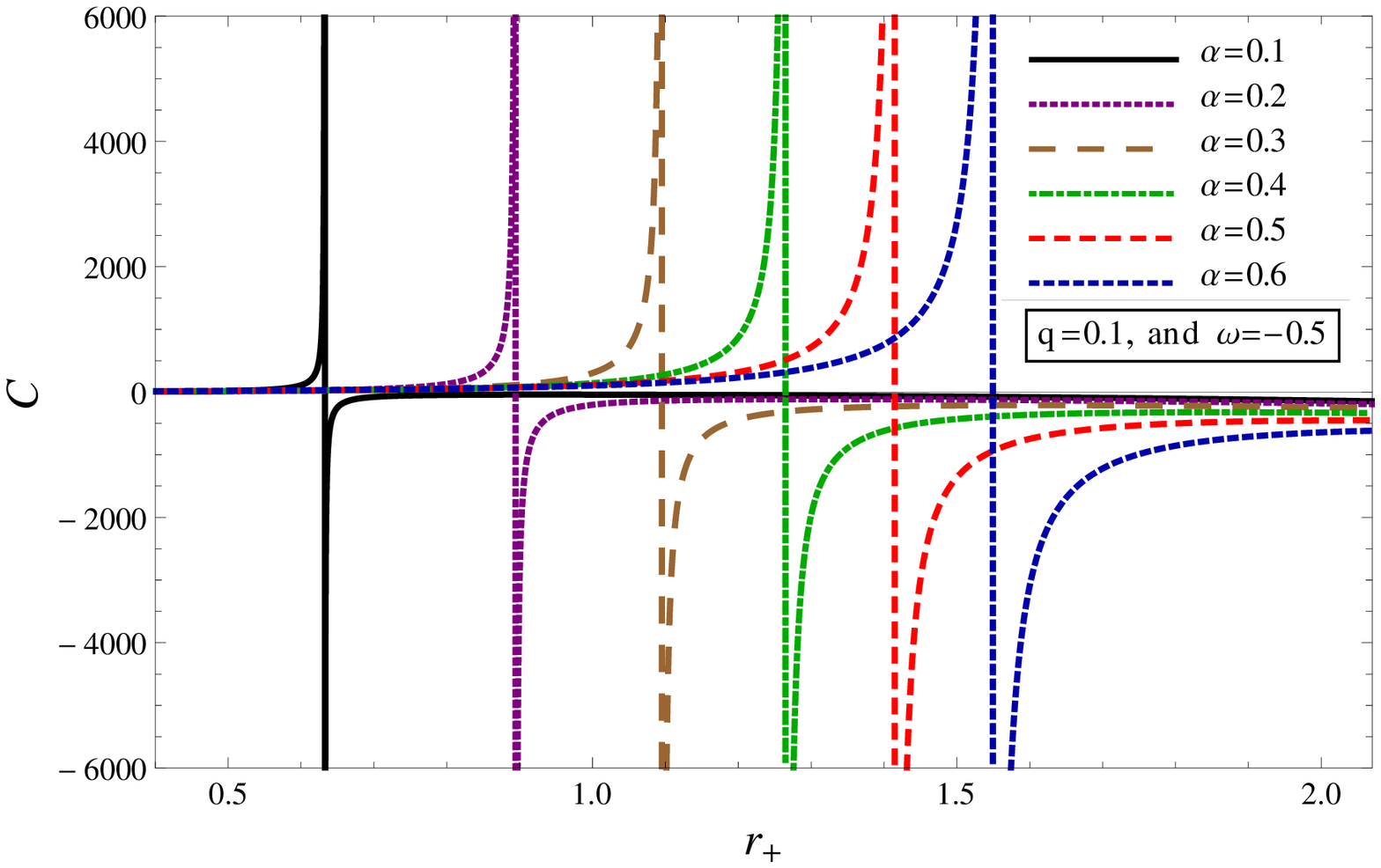}
\includegraphics[width=0.5\linewidth]{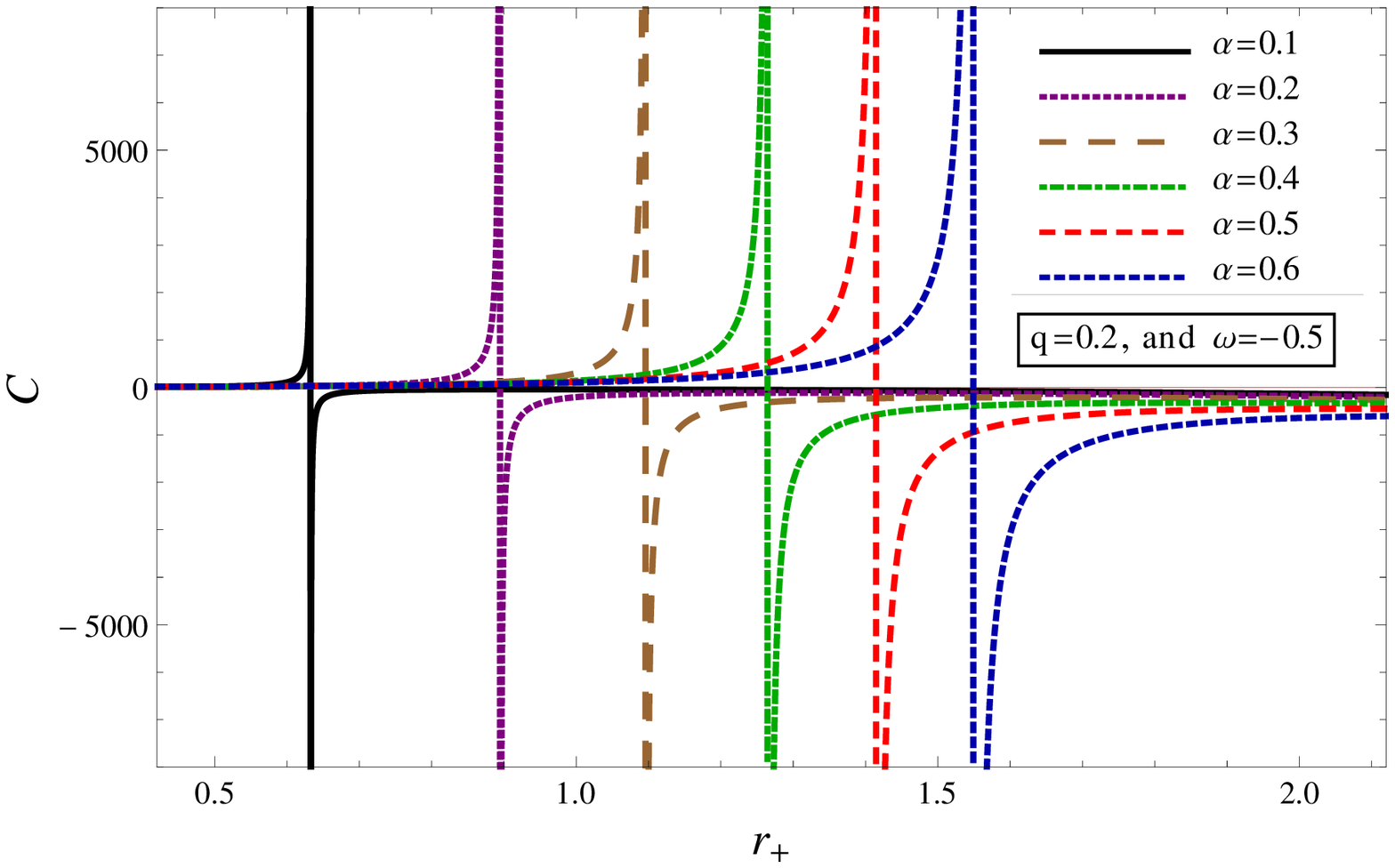}
\end{tabular}
\caption{\label{fig:egb:sh} The specific heat  $ (C_+^{\mbox{EGB}}  )$ vs horizon radius $r_+$ for 
the $ 5D $ Einstein-Gauss-Bonnet black hole surrounded by the quintessence matter for different values of 
$\omega$ and $\alpha$.}
\end{figure*}
A black hole behaves as a thermodynamical system; quantities associated with it must obey the first law 
of thermodynamics \cite{Cai:2003kt}:
\begin{equation}
dM = T_{+}dS_{+}.
\end{equation}
Hence, the entropy \cite{jdb} can be obtained from the integration 
\begin{eqnarray}\label{ent:formula}
S_{+} = \int {T_+^{-1} dM} = \int {T_+^{-1}\frac{\partial M}{\partial r_+} dr_+}.
\end{eqnarray}
Now, the entropy of the Einstein-Gauss-Bonnet gravity black holes surrounded by the quintessence matter, 
reads
\begin{eqnarray}\label{S:EGB}
S_{+}^{\mbox{EGB}} = \frac{4 \pi r_{+}^3}{3} + 16 \pi \alpha r_{+}.
\end{eqnarray}
However, it is interesting to note that the entropy of the  black hole has no effect of the 
quintessence matter parameter.

Next, let us calculate the Wald entropy for the $5D$ black hole (\ref{metric}) with $f(r)$  given by 
Eq.~(\ref{sol:egb}). Wald \cite{Wald:1993nt} showed that the black hole entropy can be calculated by
\begin{equation}\label{sw}
S_W = - \int_{\Sigma} \left(\frac{\partial \mathcal{L}}{\partial R_{abcd}}\right)\varepsilon_{ab} 
\varepsilon_{cd} dV^2_3,
\end{equation}
where $dV_3$ is the volume element on $\Sigma$ and the integral is performed on $3D$ space-like surface 
$\Sigma$. $\varepsilon_{ab}$ is the binormal vector to surface $\Sigma$ normalized as 
$\varepsilon_{ab} \varepsilon^{ab} = -2$, and the $\mathcal{L}$ is Lagrangian density as in (\ref{action}).
We note that the integrand can be calculated as
\begin{equation}\label{q}
\frac{\partial \mathcal{L}}{\partial R_{abcd}}\varepsilon_{ab} \varepsilon_{cd} = -2-
\frac{24 \alpha [1- f_{-}(r_+)]}{r_+^2}.
\end{equation}
On substituting the Eq.~(\ref{q}) into the Eq.~(\ref{sw}), one obtain Wald entropy of the $5D$ black hole (\ref{metric}) as 
\begin{eqnarray}\label{wentropy}
S_W &=& \left[1 + \frac{12 \alpha [1- f_{-}(r_+)]}{r_+^2}\right]\int_{\Sigma}  dV^2_3, \nonumber \\
 &=& \frac{4\pi r_+^3}{3}\left[1 + \frac{12 \alpha [1- f_{-}(r_+)]}{r_+^2}\right], \nonumber \\
 &=& \frac{4 \pi r_{+}^3}{3} + 16 \pi \alpha r_{+},
\end{eqnarray}
as $f_{-}(r_+)=0$ and hence  the Wald entropy Eq.~(\ref{wentropy})  has exactly same expression as obtained in Eq.~(\ref{S:EGB}). 
Furthermore, we verify that the $5D$ black hole (\ref{metric}) satisfy the first law of thermodynamics. 
The variation of Wald entropy (\ref{wentropy}) with respect to the radius $r_+$ gives
\begin{eqnarray} \label{dSw}
d S_W &=& 4\pi (r_+^2 + 4\alpha),
\end{eqnarray}
and the variation of mass (\ref{M1}) leads to 
\begin{eqnarray} \label{dM-dS}
d M^{EGB}_+  &=& 2r_+ \left(1+ \frac{2 q \omega}{r_+^{4\omega +2}}\right).
\end{eqnarray}
Hence, with the help of Eqs.~(\ref{temp1}), (\ref{dSw}) and (\ref{dM-dS}), one can conclude that
\begin{equation}
d M^{EGB}_+ = T^{EGB}_{+} d S_W. 
\end{equation}
Hence the first law of the black hole thermodynamics holds for $5D$ black hole (\ref{metric}) with $f(r)$  given by 
Eq.~(\ref{sol:egb}).

Finally, we analyze how the quintessence matter influences the thermodynamic stability of the Einstein-Gauss-Bonnet black holes. 
The heat capacity of the black hole is defined as \cite{Cai:2003kt},
\begin{equation}
\label{sh_formula}
C_{+}= \frac{\partial{M}}{\partial{T_+}}= \frac{\partial{M}}{\partial{r_+}} \frac{\partial{r_+}}{\partial{T_+}}. 
\end{equation} 
The heat capacity of Einstein-Gauss-Bonnet black hole surrounded by the quintessence matter, using Eqs.~(\ref{M1}), (\ref{temp1}), and (\ref{sh_formula}) is given by
\begin{eqnarray} \label{egb:sh}
C_{+}^{\mbox{EGB}} &=& \frac{-4\pi r_{+}^3 \left[1 + \frac{2 q\omega}{r^{4\omega +2}}\right](r_{+}^2+4\alpha)^2}{-4 \alpha r_{+}^2\left[1-\frac{2q\omega(1+4\omega)}{r_{+}^{4\omega +2}}\right]
+ r_{+}^4\left[1+\frac{2q\omega(3+4\omega)}{r^{4\omega+2}}\right]}.\nonumber\\
\end{eqnarray}
It is clear that the heat capacity depends on both the Gauss-Bonnet coefficient $\alpha$, and a quintessence matter parameter $\omega$. When $\alpha\rightarrow0$, it returns to the general relativity case. If in addition $q=0$, it becomes 
\begin{eqnarray} \label{egb:sh_lim}
C_+ &=& \frac{4\pi r_{+}^3 (r_{+}^2+4\alpha)^2}{4 \alpha r_{+}^2-r_{+}^4},
\end{eqnarray}
which is exactly same as the Einstein-Gauss-Bonnet case \cite{Sahabandu:2005ma,Ghosh:2014pga}. We again recall that for $C>0$ ($C<0$), the black hole is thermodynamically stable (unstable). It is difficult to analyze the heat capacity analytically hence, we plotted it in Fig.~\ref{fig:egb:sh}, for different values of $\alpha$ and $\omega$. Again, we note that there is a change of sign in the heat capacity around $r_C$, and $C$ is discontinuous at $r_+=r_C$. The heat capacity is positive for $r_+<r_C$ and thereby suggesting the thermodynamical stability of a black hole.  On the other hand, the black hole is unstable for $r_+>r_C$. Thus, the heat capacity of an Einstein-Gauss-Bonnet black hole, for different values of $\omega$ and $\alpha$, is positive for $r_+ < r_C$, while for $r_+>r_C$, it is negative.  The phase transition occurs from a lower mass black hole with the negative heat capacity to a higher mass black hole with positive heat capacity. 

It may be noted that the critical radius $r_C$ changes drastically in the presence of the quintessence matter, thereby affecting the thermodynamical stability. Indeed, the value of critical $r_C$ increases with the increase in the quintessence matter parameter $\omega$ for a given value of the Gauss-Bonnet coupling constant $\alpha$. Further, $r_C$ is also sensitive to the Gauss-Bonnet parameter $\alpha$ (cf. Fig.~\ref{fig:egb:sh}), and the critical parameter $r_C$ also increases with $\alpha$.

\section{Conclusion}
\label{conclusion}
The Einstein-Gauss-Bonnet theory has a number of additional nice properties than Einstein's general relativity that are not enjoyed by other higher-curvature theories. Hence, Einstein-Gauss-Bonnet theory  received significant attention, especially when finding black hole solutions. We have obtained an exact $5D$ static spherically symmetric black hole solutions to Einstein-Gauss-Bonnet gravity surrounded by the quintessence matter. We then proceeded to find exact expressions, in Einstein-Gauss-Bonnet gravity, for the thermodynamical quantities like the black hole mass, Hawking temperature, entropy, specific heat and in turn also analyzed the thermodynamical stability of black holes. It turns out that due to quintessence correction to  the black hole solution, the thermodynamical quantities are also getting corrected except for the entropy which does not depend on the background quintessence. The entropy of a black hole in Einstein-Gauss-Bonnet gravity doesn't obeys the area law.

The phase transition is characterized by the divergence of specific heat at a critical radius $r_C$ which is changing with Gauss-Bonnet parameter $\alpha$ as well as with $w$. In particular, the black hole is thermodynamically stable, with a positive heat capacity for the range $0 < r < r_C$ and unstable for $r>r_C$. It would be important to understand how these black holes with positive specific heat ($C>0$) would emerge from thermal radiation through a phase transition. We also discuss the phase transition of the black holes. The results presented here are the generalization of the previous discussions, on the Einstein-Gauss-Bonnet black hole, in a more general setting, and the possibility of a further generalization of these results to Lovelock gravity is an interesting problem for future research.

\acknowledgements
S.G.G. would like to thank SERB-DST research project Grant No. SB/S2/HEP-008/2014, and ICTP for 
Grant No. OEA-NET-76. S.D.M. acknowledges that this work is based 
upon research supported by South African Research Chair Initiative of the Department of Science and 
Technology and the National Research Foundation. M.A. acknowledges the University Grant Commission, India, 
for the financial support through the Maulana Azad National Fellowship For Minority Students scheme 
(Grant No. F1-17.1/2012-13/MANF-2012-13-MUS-RAJ-8679). We would like to thank the India-South Africa 
bilateral project Grant No. DST/INT/South Africa/P-06/2016 date: 12/07/2016, and to IUCAA, Pune for the 
hospitality, where a part of this work was done.  

\appendix

\section{Exact solutions for general relativity}
\label{sol-GR}
Making use of Eqs.~(\ref{EMT}) and (\ref{master}),  for $\alpha=0$, we obtain 
\begin{equation}\label{eehd}
r^2 f''(r) + (5+4 \omega)rf'(r)+2(2+4\omega)(f(r)-1)=0,
\end{equation}
in which a prime denotes a derivative with respect to $r$. The Eq.~(\ref{eehd}) admits a general solution
which describes a $5D$ black hole  surrounded by the quintessence matter, and the corresponding metric for the spherically symmetric takes the form
\begin{eqnarray}\label{metric5d}
ds^2 &=& -\left[1-\frac{M}{r^2}- \frac{q}{r^{4\omega +2}}\right] dt^2+ \frac{1}{\left[1-\frac{M}{r^2}- \frac{q}{r^{4\omega +2}}\right]} dr^2 \nonumber \\ 
&& + r^2 d\omega^2_3,
\end{eqnarray}
with $d\omega^2_3$ is metric on the 3-sphere. This solution for $d$-dimensional case was found in Ref. \cite{Chen:2008ra}. In order to study the general structure of the solution (\ref{metric5d}), we look for the essential singularity by calculating the Kretschmann scalar ($\mathcal{K}=R_{\mu \nu \gamma \delta} R^{\mu \nu \gamma \delta}$), that for the metric (\ref{metric}) after inserting Eq.~(\ref{metric5d}) reads
\begin{eqnarray}\label{ks}
\mathcal{K}&=& \frac{18 M}{r^8}+ \frac{ 12 (1+\omega)(4\omega +3)q M}{r^{4(\omega +2)}} + \frac{B q^2}{r^{8(\omega +1)}},
\end{eqnarray}
with 
$ B = 6\left(32\omega^4 +80\omega^3 +86 \omega^2 +42\omega +9\right)$. The Kretschmann scalar diverge as $r \rightarrow 0$, indicating the scalar polynomial or essential singularity at $r=0$ \cite{he}. The energy density ($\rho$) and the pressure ($P$) for the quintessence matter can be expressed as
\begin{equation}\label{density}
\rho = -\frac{6 q \omega}{r^{4(\omega +1)}}, \quad P = \frac{2 q \omega (1+4\omega)}{r^{4(\omega+1)}},
\end{equation} 
and
\begin{equation}
\rho + P = \frac{4 q \omega (2\omega-1)}{r^{4(\omega+1)}}.
\end{equation}
The weak energy condition is satisfied since, $\rho_q \geq 0$ and $\rho_q +P_q \geq 0$, for $-1< \omega <0$. 
When $\omega=-1$, the metric (\ref{metric5d}) takes the form of $5D$ Schwarzschild-de Sitter black hole.
The metric (\ref{metric5d}) also reduces to $5D$ Reissner-Nordstr{\"o}m black hole when $\omega=1/2$. 
The solution (\ref{metric5d}) represents a general class of static, spherically symmetric solutions 
to the Einstein's equations describing the black holes with the energy-momentum tensor that of the 
quintessence matter. The solution (\ref{metric5d}) include several known spherically symmetric solutions 
of the Einstein field equations, for instance, the $5D$ Schwarzschild solution for $q=0$, including 
its generalization to asymptotically de Sitter/Anti-de Sitter (dS/AdS) for $q(v)\neq0, \omega=-1$ 
\cite{Ghosh:2008zza}. Obviously, by proper choice of the functions $M$ and $q$, and $\omega$ parameter, 
one can generate many other solutions \cite{Ghosh:2008zza}. 
\begin{figure*}
\begin{tabular}{ c c }
\includegraphics[width=0.5\linewidth]{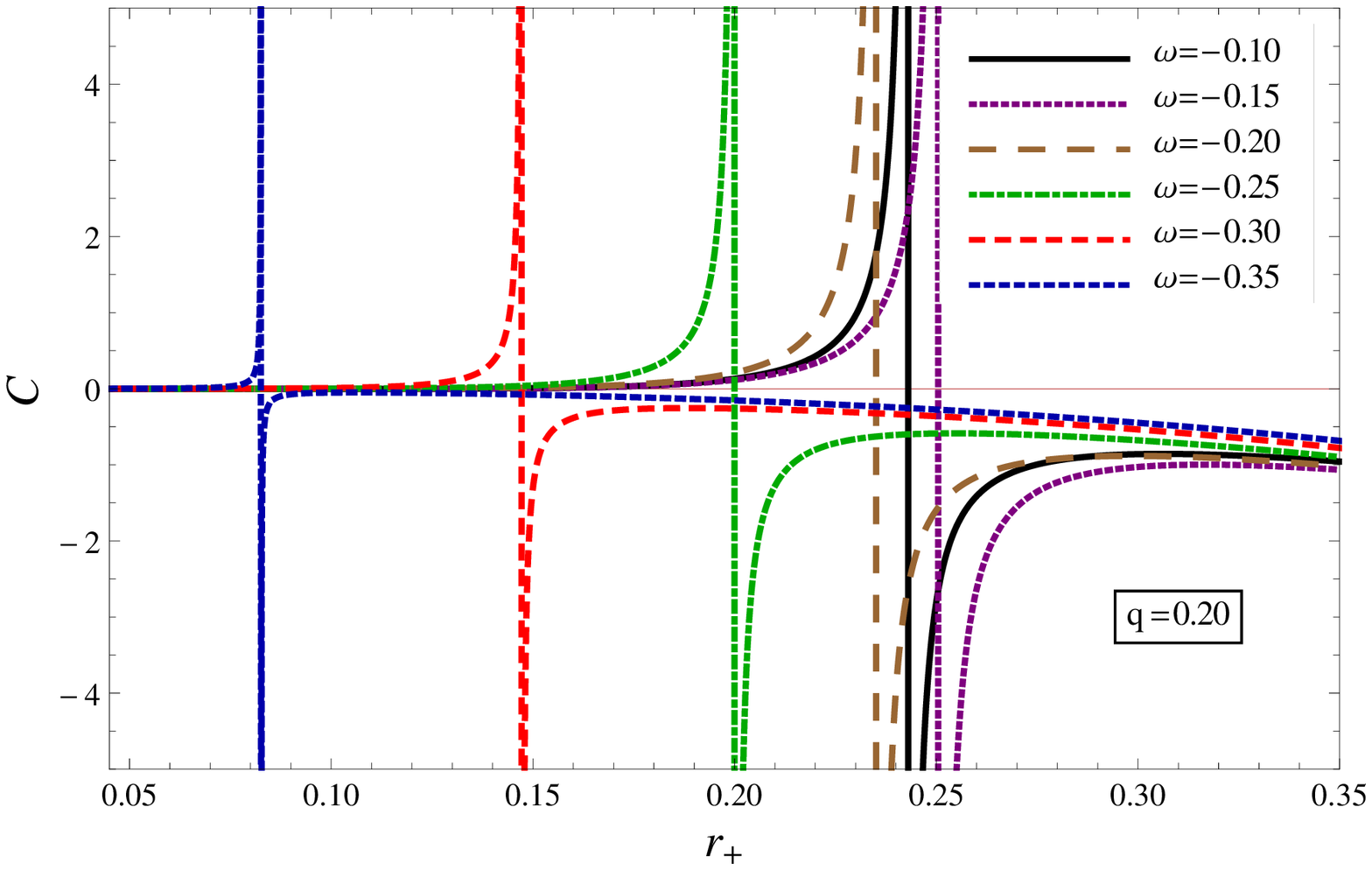}
\includegraphics[width=0.5\linewidth]{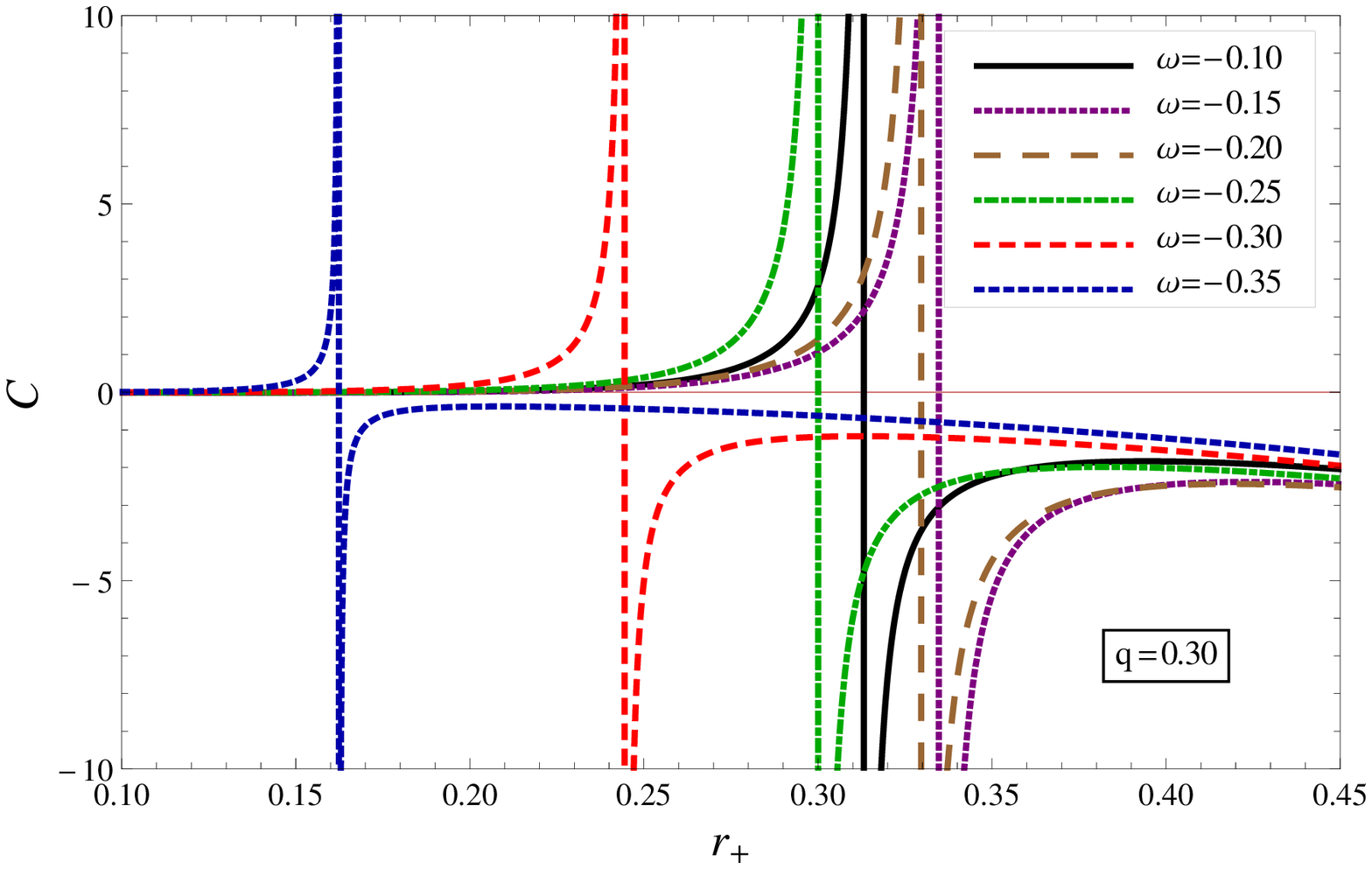}
\end{tabular}
\caption{\label{5dc1} The specific heat ($C_+$) vs horizon radius $r_+$ for the $5D$ general relativity 
black hole surrounded by the quintessence matter. The black hole thermodynamically unstable when 
$C_+<0$, and stable when $C_+>0.$}
\end{figure*}

Next, we analyze the thermodynamics of the quintessence corrected black hole given by the metric \
(\ref{metric5d}). The event horizon $r_+$ of the black hole, satisfy $g^{rr}(r_+)=0$, i.e.,
\begin{equation}
r^{4\omega+2}-M r^{4\omega} -q=0.
\end{equation}
On the other hand, the quintessence matter alone ($M=0$) has horizon placed at $r_+ =(q)^{1/4\omega+2}$. Obviously, in the limit $q \rightarrow 0$, the above solution will reduce to $5D$ general relativity black hole in which case $R=\mathcal{R}=0$. Next, we shall discuss the thermodynamics of the $5D$ black hole surrounded by the quintessence matter. We note that the gravitational mass of a black hole is determined by $g^{rr}(r_+)=0$, which, from Eq.~(\ref{metric5d}), reads
\begin{eqnarray}\label{mass}
M_{+} = r_{+}^2\left[1 - \frac{q}{r_{+}^{4\omega +2}}\right].
\end{eqnarray}
Eq.~(\ref{mass}) takes the form of the $5D$ Schwarzschild black hole mass $M=r_+^2$, when $q\rightarrow0$ \cite{Ghosh}. Accordingly, the Hawking temperature of the black hole at outer horizon $r_+$, reads
\begin{eqnarray}
T_{+} = \frac{\kappa}{2\pi}=\frac{1}{2 \pi r_{+}} \left[1+ \frac{2q\omega}{r_{+}^{4\omega +2}}\right]. \label{temp}
\end{eqnarray}
Then, we can easily see that the temperature is positive. Taking the limit $q\rightarrow 0$, we recover 
the temperature for $5D$ general relativity $T_{+} = \frac{1}{2\pi r_+}$ 
\cite{Sahabandu:2005ma,Ghosh:2014pga}, which shows that Hawking temperature diverges as $r_+ \rightarrow 0$. 
Next, we turn to calculate the Wald entropy associated with the black hole horizon from 
Eq.~(\ref{wentropy}), we arrive at  
\begin{eqnarray}\label{S:GR}
S_{W} = \frac{4\pi r_{+}^{3}}{3}. 
\end{eqnarray}
Thus, we note that the quantity $4\pi r_{+}^3 /3$ of Eq.~(\ref{S:GR}) is just an area of the black hole horizon. In proper units the Eq.~(\ref{S:GR}) may be written as $S_{+}=A/4G$ (see Appendix B). Thus,  we conclude that the entropy of the $5D$ black hole obeys the area law, and the entropy of a black hole has no influence of the quintessence matter. Next, we turn our attention to the stability of the black holes by calculating the specific heat of the black hole solution (\ref{metric5d}) and to discuss the effect of the quintessence matter. The Schwarzschild  black hole and higher-dimensional Schwarzschild-Tangherlini always have negative heat capacity, indicating thermodynamic instability of these black holes \cite{Cai:2001dz}. Inserting Eqs.~(\ref{mass}) and (\ref{temp}) in (\ref{sh_formula}), we obtain
\begin{eqnarray}\label{gr:sh}
C_{+} &=& \frac{-4 \pi r_{+}^{3}\left[ 1 + \frac{2q \omega}{r_+^{4 \omega +2}}\right]}{\left[ 1 + \frac{2q \omega(3+4 \omega)}{r_+^{4 \omega +2}}\right]}.
\end{eqnarray}
It is well known that the thermodynamic stability of the system is related to the sign of the heat capacity ($C$). If the heat capacity is positive ($C>0$), then the black hole is stable; when it is negative ($C<0$), the black hole is said to be unstable. It is clear from Eq.~(\ref{gr:sh}) that the heat capacity depends on the quintessence matter. When $q \rightarrow 0$, one gets $C=-4\pi r^3_+$, which means $5D$ general relativity black holes are thermodynamically unstable \cite{Cai:2003kt,Ghosh:2014pga}. Next, we analyze the effect of the quintessence matter on thermodynamical stability of a black hole. We plot specific heat ($C$) with radius $r$ in Fig.~\ref{5dc1} for different values of the parameter $\omega$ and $q$. It is seen that the heat capacity discontinuous at $r=r_C$, for each $\omega$, and for a given $q$. We observe that the heat capacity is $C>0$ ($C<0$) for $r_+ < r_C$ ($r_+ >r_C$). Thus, the $5D$ black hole is thermodynamically stable for $r<r_C$, as the black hole has the positive heat capacity and unstable for $r>r_C$ (cf. Fig.~\ref{5dc1}).  The black hole mass increases with increase in $r_+$. Hence, the phase transition occurs from a lower mass black hole with negative heat capacity to a higher mass black hole with positive heat capacity.

\section{Thermodynamics of d-dimensional black holes}
The metric of $d$-dimensional spherically symmetric black hole surrounded by quintessence reads \cite{Chen:2008ra}
\begin{widetext}
\begin{eqnarray}
ds^2 &=& -\left[1-\frac{16\pi G^{(d)}M}{(d-2)\Gamma^{(d-2)}c^4  r^{(d-3)}}- \frac{q}{r^{[(d-1)\omega + (d-3)]}}\right]dt^2 \nonumber \\ &&+\left[1-\frac{16\pi G^{(d)}M}{(d-2)\Gamma^{(d-2)}c^4  r^{(d-3)}}- \frac{q}{r^{[(d-1)\omega + (d-3)]}}\right]^{-1}dr^2 +r^2 d \Omega^{(d-2)}. \nonumber
\end{eqnarray}
\end{widetext}
where $ G^{(d) }$  is the $ d $-dimensional Newton constant,  $d \Omega^{(d-2)}$ is the line element on a unit  $ (d -2)  $ sphere, and $\Gamma^{(d-2)}$  is the volume of the unit  $ (d-2 )$ sphere.
The black hole has a horizon at $r=r_+$, which is solution of $f(r_+)=0$
\begin{eqnarray}
r_+^{[(d-1)\omega+d-3]}-\frac{16\pi G^{(d)}M}{(d-2)\Gamma^{(d-2)}c^4} r_+^{(d-1)\omega} -q=0. \nonumber
\end{eqnarray}
The black hole mass in terms of horizon $r_+$ reads
\begin{eqnarray}
M= \frac{(d-2)\Gamma^{(d-2)}c^4  r_+^{(d-3)}}{16\pi G^{(d)}}\left[1-\frac{q}{r_+^{[(d-1)\omega+ (d-3)]}}\right]. \nonumber
\end{eqnarray}
The Hawking temperature is
\begin{eqnarray}
T_+= \frac{\hbar c(d-3)}{4\pi r_+}\left[ 1-\frac{(d-1)}{(d-3)}\frac{q\omega}{r_+^{[(d-1)\omega+ (d-3)]}} \right]. \nonumber
\end{eqnarray}
By applying the Wald entropy formula (\ref{wentropy}), one can deduce the general expression for the 
entropy as 
\begin{equation}
S_{W}= \frac{1}{4\hbar G^{(d)}} c^3 \Gamma^{(d-2)} r^{(d-2)}_{+} = \frac{\textsf{A}}{4 G^{(d)}},\nonumber
\end{equation}
and it satisfy the area law.

\end{document}